\documentclass[aps,prl,twocolumn]{revtex4}

\usepackage{color}
\usepackage{epsfig,pstricks}

\usepackage{amsmath,amssymb}

\def\CC{{\rm\kern.24em \vrule width.04em height1.46ex depth-.07ex
\kern-.30em C}}
\def\P{{\rm I\kern-.25em P}}
\def\NN{{\rm I\kern-.25em N}}
\def\RR{{\rm
         \vrule width.04em height1.58ex depth-.0ex
         \kern-.04em R}}
\def\id{{\rm 1\kern-.22em l}}
\def\ZZ{{\sf Z\kern-.44em Z}}
\def\tr{{\rm tr}\;}
\newtheorem{pdef}{Definition}[section]

\newenvironment{eqblock}[2]{\beq\label{#2}\begin{array}{#1}}{\end{array}
                                \eeq}
\newenvironment{neqblock}[1]{\[\begin{array}{#1}}{\end{array}\]}

\newcommand{\beqb}{\begin{eqblock}}
\newcommand{\eeqb}{\end{eqblock}} 
\newcommand{\nbeqb}{\begin{neqblock}}
\newcommand{\neeqb}{\end{neqblock}} 

\newcommand{\beq}{\begin{equation}}
\newcommand{\beqa}{\begin{eqnarray}}
\newcommand{\eeq}{\end{equation}}
\newcommand{\eeqa}{\end{eqnarray}}
\newcommand{\nbeqa}{\begin{eqnarray*}}
\newcommand{\neeqa}{\end{eqnarray*}}

\newcommand{\bra}[1]{\langle #1 |}
\newcommand{\ket}[1]{| #1 \rangle}
\newcommand{\ketbra}[2]{\ensuremath{| #1 \rangle \langle #2 |}}
\newcommand{\braket}[2]{\langle #1 | #2 \rangle}

\def\DJo{$\;$\kern-.4em \hbox{D\kern-.8em\raise.15ex\hbox{--}\kern.35em okovi\'c}}

\begin{document}

\title{The fourtangle in the transverse XY model}
\author{Andreas Osterloh}
\affiliation{Institut f\"ur Theoretische Physik, 
         Universit\"at Duisburg-Essen, D-47048 Duisburg, Germany.}
\email{andreas.osterloh@uni-due.de}
\author{Ralf Sch\"utzhold}
\affiliation{Institut f\"ur Theoretische Physik, 
         Universit\"at Duisburg-Essen, D-47048 Duisburg, Germany.}
\email{ralf.schuetzhold@uni-due.de}
\begin{abstract}
We analyze the entanglement measure $C_4$ for mixed states in general and for the transverse XY model.
We come to 
the conclusion that it cannot serve alone for guaranteeing an entanglement of $GHZ_4$-type.
The genuine negativity calculated in Ref.~\cite{Hofmann14} isn't sufficient for that either and 
some additional measure of entanglement must be considered. 
In particular we study the transverse XY-model and find a non-zero $C_4$ measure which is of the same 
order of magnitude than the genuine negativity.
Furthermore, we observe a feature in the $C_4$ values that resembles a
destructive interference with the underlying concurrence. 
\end{abstract}

\maketitle

\section{Introduction}
Entanglement is a resource in physics and therefore needs to be quantified 
and to be better understood. For this sake it is of major importance to 
quantify and classify entanglement in laboratory systems, hence for mixed states. 
In the year 2002,
the works \cite{OstNat,Osborne02} have initiated an avalanche of workanalysis into this 
direction in the following decade. Particular importance was drawn to the
Coffman-Kundu-Wootters (CKW) inequality\cite{Coffman00} 
which connected the total entanglement detectable - the tangle: a quantity that originates in 
the single site reduced density matrix - with something not encoded in
the entanglement of pairs, as measured by the concurrence. The difference 
of the tangle and the sum of the concurrences squared was henceforth interpreted as residual entanglement.
The residual entanglement vanishes for the $W$-states an is maximal for any maximally entangled state
with respect to the group $SL(2)$\cite{OS04,OS05,OS09}. 
The CKW conjecture could be proved in 2006 by Osborne and Verstraete in
\cite{Osborne06}. As a matter of fact, the residual tangle was shown to be dominant
over the concurrence in the transverse XY model\cite{amico04}.
This means that most of the present quantum correlations
around its quantum phase transitions must come from genuine 
multipartite entanglement in the spirit of ref.~\cite{OS04}. 
Since then, there have been only few recent trials
of looking into that direction\cite{Hiesmayr13,Hofmann14}. 
Here, we will follow this road with an entanglement measure, 
which is the 4-Tangle $C_4$, the 4-particle generalization of the concurrence $C_2$. 
It has been introduced for pure states in Ref.~\cite{Wong00} and its convex roof extension is 
due to Uhlmann\cite{Uhlmann}. This choice is rather obviously taken with regard to its simple 
handling and in that it only detects GHZ-type of states and products of Bell-states\cite{DoOs08}.
So besides a possible bipartite part, any further entanglement detected by it will be of GHZ-type.
The genuine negativity\cite{Jungnitsch11} is only detecting states that are not biseparable.
Therefore it will not detect products of Bell-states, but it will detect W-type of states.
So, when looking at $C_4$ in parallel to the genuine negativity only, 
we cannot certify GHZ-type entanglement, since the negativity may only detect entanglement of $W$-kind,
and $C_4$ detects also mixtures of $W$-states and biseparable products of Bell-states. The results will 
therefor be at most a hint towards GHZ-entanglement in this model.

This work is laid out as follows:
we begin with a study of $C_4$ for states in general in the next section. 
Next we analyze this quantity for the transverse Ising model 
followed by the $XY$-model.
The conclusions are drawn and an outlook on possible future directions is given in the last section.

\section{The entanglement measure $C_4$}\label{4-tangle}

We highlight on an $SL$ invariant measure of entanglement,
the $4$-tangle $C_4[\psi]:=|\bra{\psi^*}\sigma_y^{\otimes 4}\ket{\psi}|$ in this work. 
Whereas this measure cannot distinguish
between entanglement that is carried by the states like 
$GHZ_4=\frac{1}{\sqrt{2}}(\ket{0000}+\ket{1111})$ from that carried by 
products of Bell states, it will not detect any entanglement supported 
by $W$ type of states\cite{OS04,OS05,DoOs08}. These
globally entangled states (but not genuinely multipartite entangled states, following the notion in
Ref.~\cite{OS05}) are detected exclusively by the $SL$ invariant 
entanglement measure on two qubits, which is the concurrence\cite{Hill97,Wootters98}.
W-states carries therfore only entangled in pairs of sites.
It's conves-roof extension is calculated for mixed states in the following way\cite{Uhlmann00}
\beqa
\Sigma_4&:=&\sigma_y\otimes\sigma_y\otimes\sigma_y\otimes\sigma_y \\
R&:=&\sqrt{\rho}\,\Sigma_4 \rho^* \Sigma_4 \sqrt{\rho}\\
C_4[\rho]&=& 2\lambda_{\rm max}-\tr \sqrt{R}
\eeqa 
where $\lambda_{\rm max}$ is the maximal eigenvalue of the non-negative operator $\sqrt{R}$.
\begin{figure}
\centering
  \includegraphics[width=.9\linewidth]{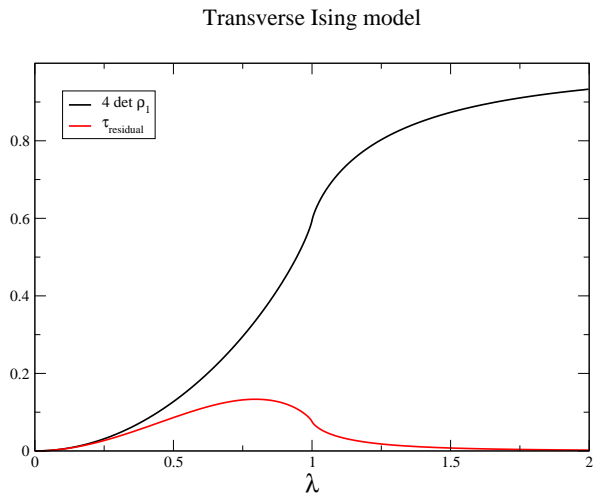}
  \caption{The solid black line is the 1-tangle, $4\det \rho_1$, and the red solid line 
$\sum_d C^2_d(\lambda,1)$. 
The difference of both is the ``residual tangle'' in the chain, which consists of 
multipartite entanglement beyond two sites.\\}
  \label{ResidualTangle}
\end{figure}
At first, we briefly analyse the $4$-tangle for mixed states. 

\begin{figure}
  \centering
  \includegraphics[width=.9\linewidth]{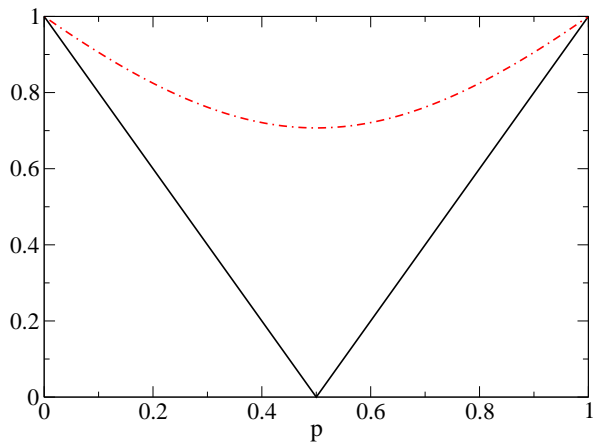}
  \caption{Here, the value of $C_4$ for $\rho=p\ketbra{GHZ_4}{GHZ_4}
          +(1-p) \ketbra{Bell\otimes Bell}{Bell\otimes Bell} $ 
is shown for $\ket{GHZ_4}=(\ket{1111}+\ket{0000})/\sqrt{2}$ and 
$\ket{Bell\otimes Bell}$ is one
of the two states $(\ket{11}+\ket{00})/\sqrt{2}$ (red dash-dotted curve) and $(\ket{11}+i \ket{00})/\sqrt{2}$ 
(black solid line).}
  \label{GHZ-BellBell}
\end{figure}

Since the $4$-tangle of a tensor product of Bell states is also maximal as for $GHZ$ states, it
is not surprising that for the state $\rho(p)=p \ketbra{GHZ_4}{GHZ_4} 
+ (1-p) \ketbra{\psi_B\otimes\phi_B}{\psi_B\otimes\phi_B}$,
with $\psi_B,\phi_B\in (\ket{\sigma,\sigma}\pm \ket{\overline{\sigma},\overline{\sigma}})/\sqrt{2}$
and $\braket{\sigma}{\overline{\sigma}}=0$, the entanglement classes interfer such that $C_4[\rho(\frac{1}{2})]$
assumes its minimal value at zero if the states are orthogonal to each other
and it can take decreasing values from at most $1/\sqrt{2}$ down to $0$ if $\ket{\sigma,\sigma}$ 
is non-orthogonal to $GHZ_4$ (see figure \ref{GHZ-BellBell}). 
Therefore we have two different classes of entanglement 
which interfere - genuinely entangled GHZ states and biseparabel products of Bell states.
The contrary holds for admixtures of a $W_4$ state;
\begin{figure}
\centering
  \includegraphics[width=.75\linewidth]{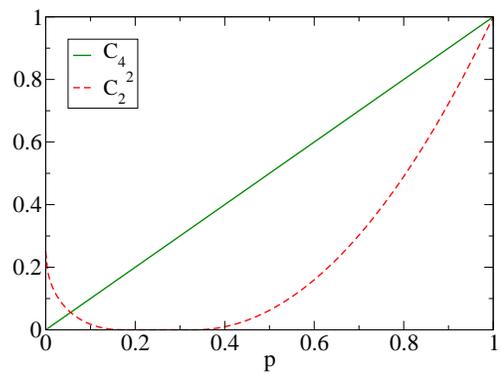}
  \caption{$C_4$ and $C_{2;1,2}C_{2;3,4}$ are shown for 
$\rho(p)=p \ketbra{\Phi^+\otimes\Phi^+}{\Phi^+\otimes\Phi^+}+(1-p) \ketbra{W}{W}$. 
Here, $C_{2;i,j}$ is the concurrence of the reduced density matrix of the sites $i$ and $j$. 
It is clearly seen that the factorizing property of $C_4$ into the concurrences for pure states 
doesn't mean 
that it factorizes also for mixed states. Whereas $C_4$ linearly decreases, $C_{2;1,2}C_{2;3,4}$ has two
distinct zeros at $p_1\sim 0.1716$ and $p_2=1/3$. Even if the square root is taken from 
the concurrences, this would mean only to replace the red curves by straight lines.}
  \label{C4-C2sqr.BellBell-W}
\end{figure}
it does not lead to an interfering behaviour
as in the case of three qubits\cite{LOSU} in that it linearly grows in $p$ for 
$\rho(p)=p\ketbra{GHZ_4}{GHZ_4} + (1-p) \ketbra{W_4}{W_4}$ or 
$\rho(p)=p \ketbra{\Phi^+\otimes\Phi^+}{\Phi^+\otimes\Phi^+}+(1-p) \ketbra{W}{W}$ from $0$ to $1$. 
In contrast, it will of course 
influence the concurrence when tracing out arbitrary two qubits, as is shown 
in fig. \ref{C4-C2sqr.BellBell-W}. We want to emphasize here that one can not infer from
the relation of $C_4$ and corresponding concurrences anything about 
the entanglement type participating in the state at hand. Here we even have a whole interval
where $C_4$ is positive and the corresponding concurrences vanish.

\begin{figure}
 \centering
  \includegraphics[width=.9\linewidth]{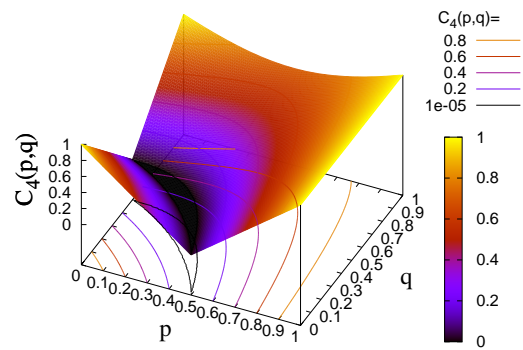}
  \caption{Here it is seen, how the admixture of an additional Bell state influences the result;
$C_4$ becomes precisely zero. The density matrix is taken to be
$\rho=p\ket{GHZ_4}+(1-p)(q\ket{\Phi^-\otimes\Phi^-} + (1-q)\ket{\Psi^-\otimes\Psi^-})$, with 
$\ket{\Phi^{\pm}}=(\ket{11}\pm\ket{00})/\sqrt{2}$, and $\ket{\Psi^{\pm}}=(\ket{10}\pm\ket{01})/\sqrt{2}$}
  \label{GHZ-BellBell-BellBell}
\end{figure}

\begin{figure}
\centering
  \includegraphics[width=.9\linewidth]{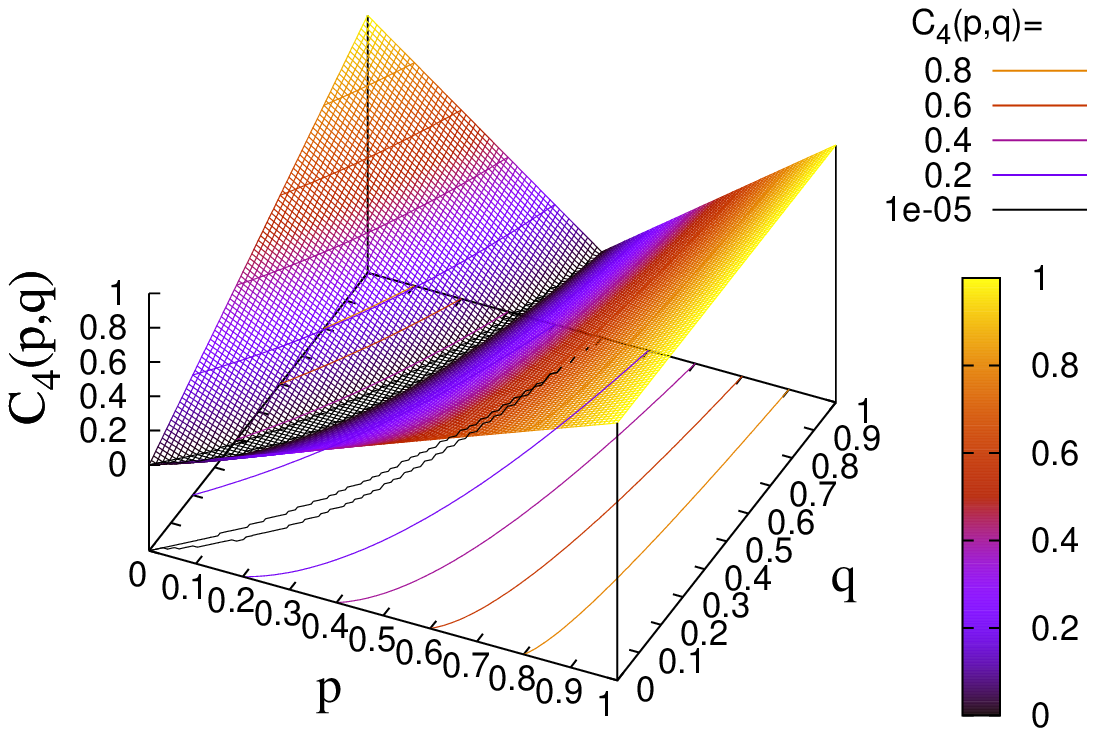}
  \caption{Here the value of $C_4$ is shown for the rank $3$ density matrix
$\rho=p\ketbra{GHZ'_4}{GHZ'_4}+(1-p)(q \ketbra{Bell\otimes Bell}{Bell\otimes Bell}
 +(1-q)\ketbra{W_4}{W_4})$ with
$\ket{GHZ'_4}=(\ket{0100}+\ket{1011})/\sqrt{2}$, $\ket{Bell\otimes Bell}=\Phi^+\otimes\Psi^-$ with 
$\Phi^+$ and $\Psi^-$ as defined in figure \ref{GHZ-BellBell-BellBell}, 
and $\ket{W_4}=(\ket{1000+\ket{0100}+\ket{0010}+\ket{0001})/2}$.}
  \label{GHZ-BellBell-W}
\end{figure}

\begin{figure}
  \centering
  \includegraphics[width=.9\linewidth]{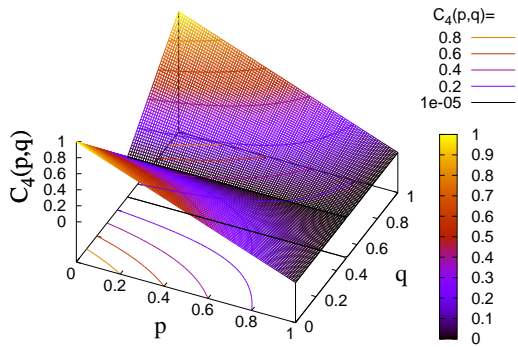}
  \caption{The value of $C_4$ for $\rho=p\ketbra{W_4}{W_4}+(1-p)
(q \ketbra{Bell_1\otimes Bell_1}{Bell_1\otimes Bell_1}
             +(1-q) \ketbra{Bell_2\otimes Bell_2}{Bell_2\otimes Bell_2} ) $ 
is shown for $\ket{W_4}=(\ket{1000}+\ket{0100}+\ket{0010}+\ket{0001})/\sqrt{2}$ and 
$\ket{Bell_i}=(\ket{11}+(-1)^i\ket{00})/\sqrt{2}$. It is zero only on the centerline between 
both Bell states and for the $W_4$ state at $p=1$.}
  \label{C4-W-BellBells}
\end{figure}

\begin{figure}
 \centering
  \includegraphics[width=.9\linewidth]{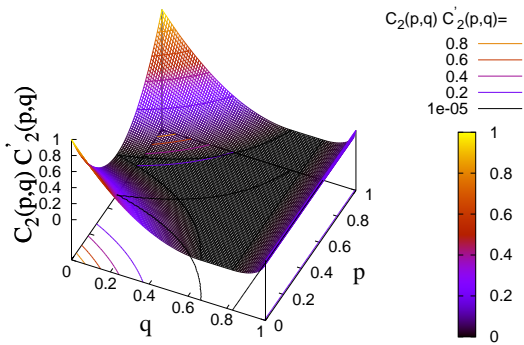}
  \caption{The product of two concurrences are shown for the same mixed state as 
           in fig. \ref{C4-W-BellBells}. The product of the two concurrences is zero
           on a whole region, which has a chock-like form. }
  \label{C2sqr-WBells}
\end{figure}

We now discuss rank three states. There are several interesting cases for the admixtures 
of $GHZ_4$ states, product of Bell states, and $W_4$ states. 
For $GHZ_4-Bell\otimes Bell-W_4$ mixtures and mixtures of $GHZ_4$ and two different products
of Bell-states, there appears a whole regions where $C_4$ is zero
(see figs. \ref{GHZ-BellBell-BellBell} and \ref{GHZ-BellBell-W}). 
But it is unclear to assign which of the two classes contributed mainly to the state. 
This becomes particularily clear when no genuinely entangled state is in
the optimal decomposition as it is shown in figs. \ref{C4-W-BellBells} and \ref{C2sqr-WBells}. 
Also in this case, as for rank two density matrices, a zero in the product of the two concurrences 
in a particular $2-2$ bipartition, which is satisfied for an almost chock-like range, 
does not mean that necessarily $C_4=0$ (which is only satisfied precisely on the centerline
of the two Bell states), as one could erroneously
conclude from its decomposition of $C_4$ for pure states into a product of any two concurrences.
The raise for the product of the concurrences from $q\sim 0.83$ quadratically to $0.25$ at $q=1$
is merely due to the $W$-state. 
For the mixture of two $GHZ$ states and the $W$ state we have the same situation as in 
fig. \ref{C4-W-BellBells}; the only difference being that the product of the concurrences 
is always zero except of its quadratic raise  
from $q\sim 0.83$ to $0.25$ at $q=1$ (see fig. \ref{C2sqr-WBells}).
We thus cannot learn from $C_4$ alone about the nature of entanglement of the state.
 
With results as those from the PPT-criterion\cite{Hofmann14} we can at best conclude that the state 
contains entanglement wich in not biseparable in a region where the 
``genuine multipartite negativity''\cite{Hofmann14} $N_\rho$ is non-zero.
This also includes mixtures of the $W$ state and any biseparable Bell-products.

\section{The transverse Ising model} \label{Sec:Ising}

Next, we analyze the spin-1/2 Ising model, which is given through the Hamiltonian
\beq\label{Ising}
H=-\lambda\sum_i S_i^x S_i^x
-\sum_i S_i^z
\eeq
This model has a second order phase transition 
from antiferromagnetism at $\lambda<-1$ via the paramagnetic phase 
at $|\lambda|<1$ 
to ferromagnetism at $\lambda>1$.
Since $C_d(\lambda,1)$ vanishes
for distances $d>2$, the monogamy relation is easily obtained\cite{amico04}, 
demonstrating that the essential entanglement in the tansverse Ising model
must be of some multipartite type (see fig. \ref{ResidualTangle}).
Of what type however has never been investigated and even the recent contributions
\cite{Hiesmayr13} and \cite{Hofmann14}) do not distinguish $W$ from $GHZ$ entanglement.
Recent discoveries would render 
this however a feasible task\cite{Eltschka2012,Siewert2012}.
We nevertheless analyze the entanglement measure $C_4$ for this model and compare with   
the results from Ref.~\cite{Hofmann14}. 

We introduce at first our notation. We write $C_4(n_1,n_2,n_3)$, 
where the numbers $n_i$ indicate how far away to the right is the next neighbor. 
$C_4(1,1,1)$ hence means that all neighbors are nearest neighbors with a distance of $1$. 

We want to highlight here that whenever the state would become a tensor product (examples are usually
states with distances $(i,n,j)$ when $n\to\infty$)
of two two-site matrices, then the optimal decomposition to the concurrences become a decomposition
of $C_4$, and therefore $C_4$ is upper bounded by the product of the concurrences.
Therefore, we take the two major concurrences as a way to confront the curves for $C_4(i,n,j)$ with.
This means that we confront $C^2_2(1)$ with $C_4(1,n,1)$, and $C_2(1)C_2(2)$ with $C_4(1,n,2)$.
Since the concurrence decays with the distance of the constituents, the committed error will be 
almost negligible.

\begin{figure}
\centering
  \includegraphics[width=.75\linewidth]{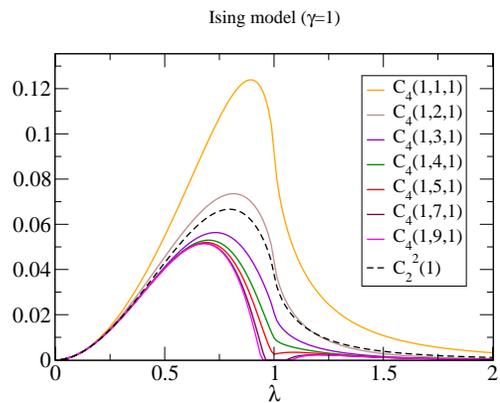}
  \caption{We show various graphs of $C_4(1,n,1)$ for $n=1$ to $9$. It is seen that a gap occurs for 
$n\geq 7$ around the critical point $\lambda_c=1$. In this region of vanishing $C_4$, the optimal decomposition 
states must be made of states from the null-cone of $C_4$, including $W$-, Cluster- and 
$X$-type of states, which all contribute to the PPT-criterion\cite{Hofmann14}.
As a comparison we also print $C_2^2(1)$, which would be an upper bound to $C_4$, 
if the state would be a tensor product. For $n\geq 3$ our results are at least compatible with that hypothesis.}
  \label{C4-1n1}
\end{figure}

We start our discussion with $C_4(1,n,1)$. Observing that the nearest neighbor concurrence 
is non-zero and assuming that the density matrix be a tensor product for $n\to\infty$, we deduce 
that the expected result would be upper bounded by the square of the nearest neighbor concurrence,
when the state is assumed to be a tensor product.
Whereas this is not true for $n=2,3$ it begins to be satisfied for growing $n$, where a gap occurs
(sometimes called in the literature ``sudden death'' and ``sudden revival'' of entanglement) 
around the critical point $\lambda_c=1$ (see fig. \ref{C4-1n1}). Since the state should 
become a tensor product only for $n\to\infty$, the results are not violating this working hypothesis
of earlier work. That it is not satisfied for $n=2,3$ is not so surprising.
$C_4$ being zero means that the density matrix can be decomposed in this region
into states exclusively from the null-cone of $C_4$, that means any of the states is of the $GHZ$-type
or a tensor product of Bell-like states in whatsoever bipartitions of the four-site subsystem.
This does not mean, however, that the decomposition could not be genuinely multipartite entangled, 
since it includes also the genuinely 4-partite entangled Cluster states and $X$-states\cite{OS05},
since these have a different state length of $4$ and $6$, respectively\cite{OS09}. 
It also includes $W$ type of states as a possibility, which sometimes are also termed as being 
``genuinely multipartite entangled''.
Within the language of this paper\cite{OS04,OS05,OS09}, the $W$-state is however only a not bipartitely 
distributed two-site entangled state, whose entanglement is solely given by the concurrences; 
its residual tangle is precisely zero\cite{Coffman00}.
When confronting this with the results of ref. ~\cite{Hofmann14} we 
find that 
the state could of course contain GHZ entanglement, but it could consist also of $W$-states and a 
bipartite product of Bell-states, as seen in fig. \ref{C4-W-BellBells}. 
In addition, the optimal decompositions for $C_4$ and the genuine negativity
could be different, a phenomenon that occurred e.g. in Ref.~\cite{LOSU} and created some ambiguity
in the types of entanglement that may enter a decomposition.
We want to mention here that for configurations $(1,n,1)$ and $n\geq 3$ the genuine negativity is zero.
Hence, there the entanglement should be made out of biseparable products of Bell-states there.
Our results go conform with the genuine negativity being zero in these instances.
\begin{figure}
\centering
  \includegraphics[width=.75\linewidth]{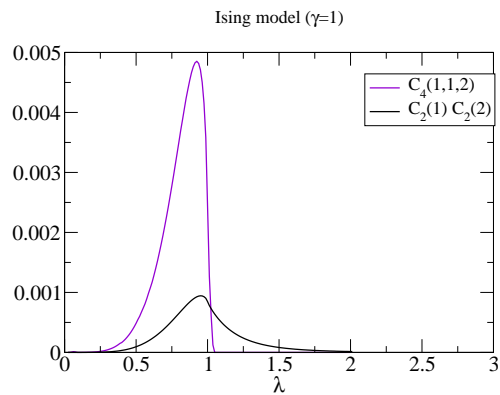}
  \caption{The figure shows $C_4(1,1,2)$ together with $C_2(1)C_2(2)$. It is the only non-vanishing
$C_4(1,n,2)$ that exists.}
  \label{C4-112}
\end{figure}

The same argument would apply to $C_4(2,n,2)$ and $C_4(1,n,2)$ or 
equivalently $C_4(2,n,1)$, but in these cases $C_4$ always turns out to be zero, 
except for $C_4(1,1,2)$, which we show in fig. \ref{C4-112}.
It is important to mention here, that $C_4(2,n,2)$ being zero does not violate the working hypothesis
that this state roughly becomes a tensor product with growing $n$, either. 
Here, the states in the optimal decomposition are in the null-cone of $C_4$ for all $C_4(2,n,2)$,
and for $n>2$ in $C_4(1,n,2)$, 
whereas there is still a possibility for the GHZ-state left to support the entanglement 
as long as $C_4$ is zero (see for instance finite regions with $C_4=0$ in
figs. \ref{GHZ-BellBell-BellBell} and \ref{GHZ-BellBell-W}).

\section{The transverse XY-model}\label{XY-model}

The Hamiltonian is
\beq\label{XY}
H=-\lambda\sum_i \left( \frac{1+\gamma}{2} S_i^x S_i^x + \frac{1-\gamma}{2} S_i^y S_i^y\right)
-\sum_i S_i^z
\eeq
and except for $\gamma=0$, the model is in 
the same universality class than the transverse Ising model. When going towards 
the isotropic model at $\gamma=0$, the range $R$ of the concurrence $C_2(n)$
grows roughly as $R\propto \gamma^{-1}$ for the critical value of the 
parameter $\lambda_c$ (see the scaling . 
The model hence merges from the Ising case for the anisotropy 
parameter $\gamma=1$ to the isotropic model for $\gamma=0$.
It has a factorizing point 
$\lambda_f=(1-\gamma^2)^{-\frac{1}{2}}$ where the ground state is a 
tensor product\cite{Kurmann,Illuminati-FF09}.
We will study more in detail the behavior of $C_4$ in the anisotropic model.

\begin{figure}
\centering
  \includegraphics[width=.75\linewidth]{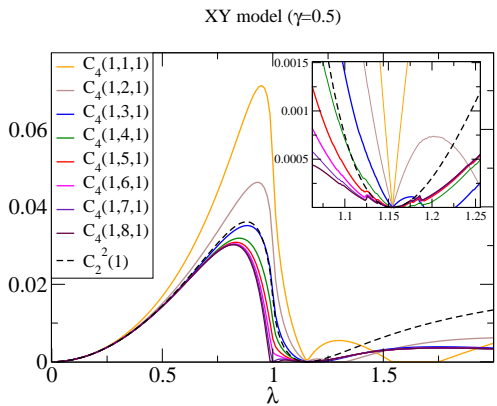}
  \caption{Various curves are plotted for $C_4(1,n,1)$ at the value $\gamma=0.5$. 
It drops down to zero around the critical point $\lambda_c=1$ as for the transverse Ising model.
This remarkable feature is not seen around the factorising point $\lambda_f=(1-\gamma^2)^{-\frac{1}{2}}$,
where the ground state is an exact site-wise tensor product. Here, $C_4(1,n,1)$ has to approach zero
at least as quickly as $C_2(1)^2$, if the state is to a good approximation a tensor product.
It, however, tends to lie a bit above $C_2(1)^2$ for $\lambda \lesssim \lambda_f$. It is seen that it 
is, however, upper bounded by $C^2_2(1)$ up to a value of $\lambda=1.125$ for $n\geq 5$. 
$C_4(1,n,1)$ is upper bounded by $C_2(1)^2$ above the factorizing field. This is seen in the inset. 
\\
For $\lambda=2$ and $n\geq 2$ they all have values of about $0.0033$.}
  \label{C4-1n1-gamma0.5}
\end{figure}

\begin{figure}
\centering
  \includegraphics[width=.75\linewidth]{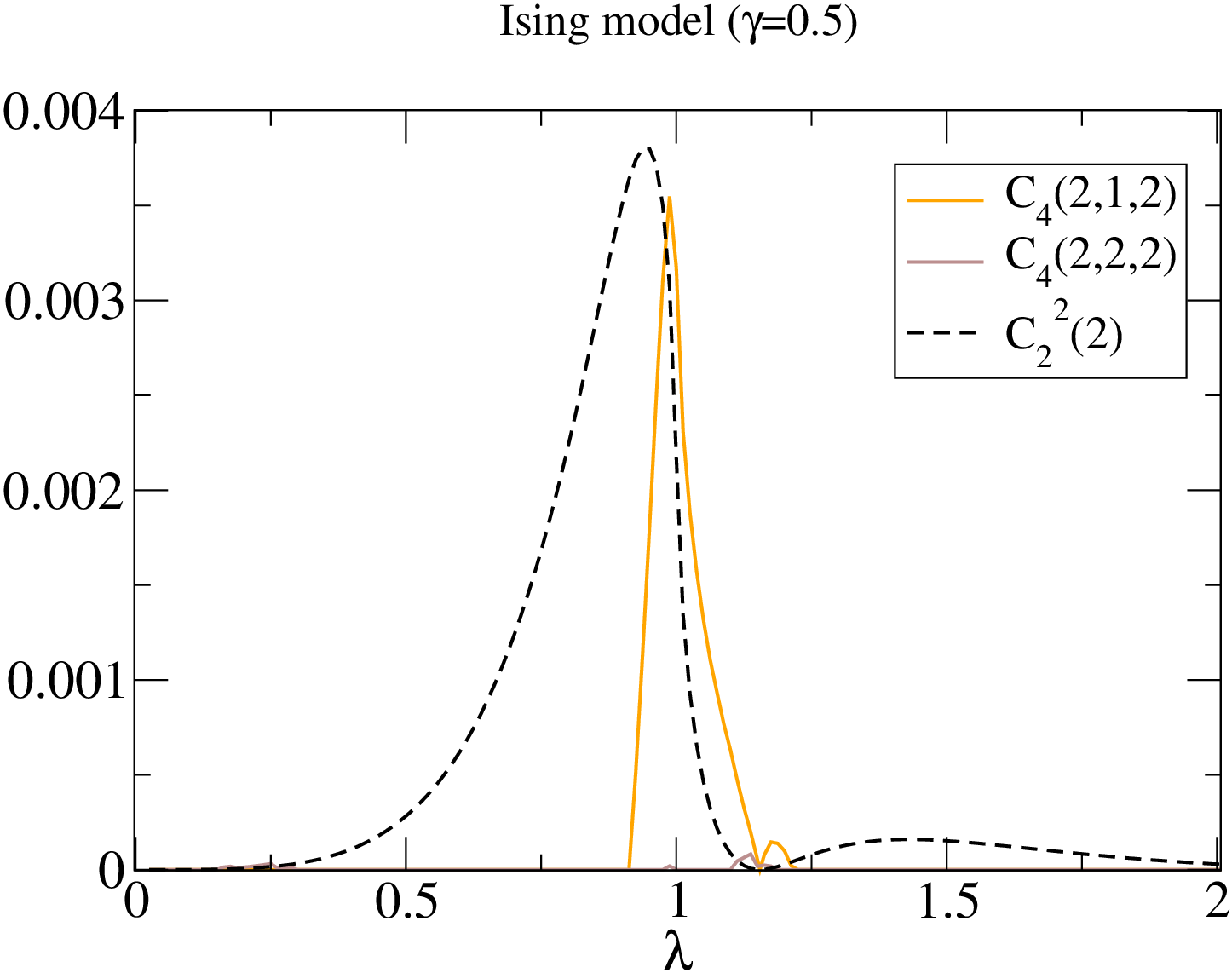}
  \caption{$C_4(2,n,2)$ is compared with $C_2^2(2)$.}
  \label{C4-2n2-gamma0.5}
\end{figure}

At first we observe that the $C_4(1,n,1)$-plots are quite similar to the
ones for the transverse Ising model except for the factorizing point, 
where every measure of entanglement vanishes. 
In particular, as far as the working hypothesis of earlier work is concerned,
$C_4(1,n,1)$ becomes upper bounded by $C_2(1)^2$ for sufficiently large $n$
(see fig. ~\ref{C4-1n1-gamma0.5}).
Besides the apparent tendency that the critical point is spared as $n$ grows,
this is not observed around the factorizing point $\lambda_f=(1-\gamma^2)^{-1/2}$.
Here, the ground state of the chain is compatible with the necessary condition
that $ C_4(1,n,1)$ be smaller than $C_2(1)^2$ for $\lambda \geq \lambda_f$; for 
$\lambda \leq \lambda_f$ this condition is violated. For $n\geq 5$ and $\lambda \leq 1.125$
it is satisfied again
(see inset of fig. \ref{C4-1n1-gamma0.5}).
$C_4(2,n,2)$ takes only a considerable part for sufficiently small $n$. 
Therefore we print it only for the values $n=1,2$ in  figure ~\ref{C4-2n2-gamma0.5}
and compare it again with the 
concurrence squared $C_2(2)^2$. That $C_4(2,n,2)$ is larger than $C_2(2)^2$ 
in a wide region for $n=1,2$ only tells
that the state has not a product form, hence it could be otherwise entangled; 
for higher values of $n$ however, we have $C_4(2,n,2)\leq C_2^2(2)$ and 
therefore the state satisfies the condition for being (roughly) a product in these cases.

\begin{figure}
\centering
  \includegraphics[width=.8\linewidth]{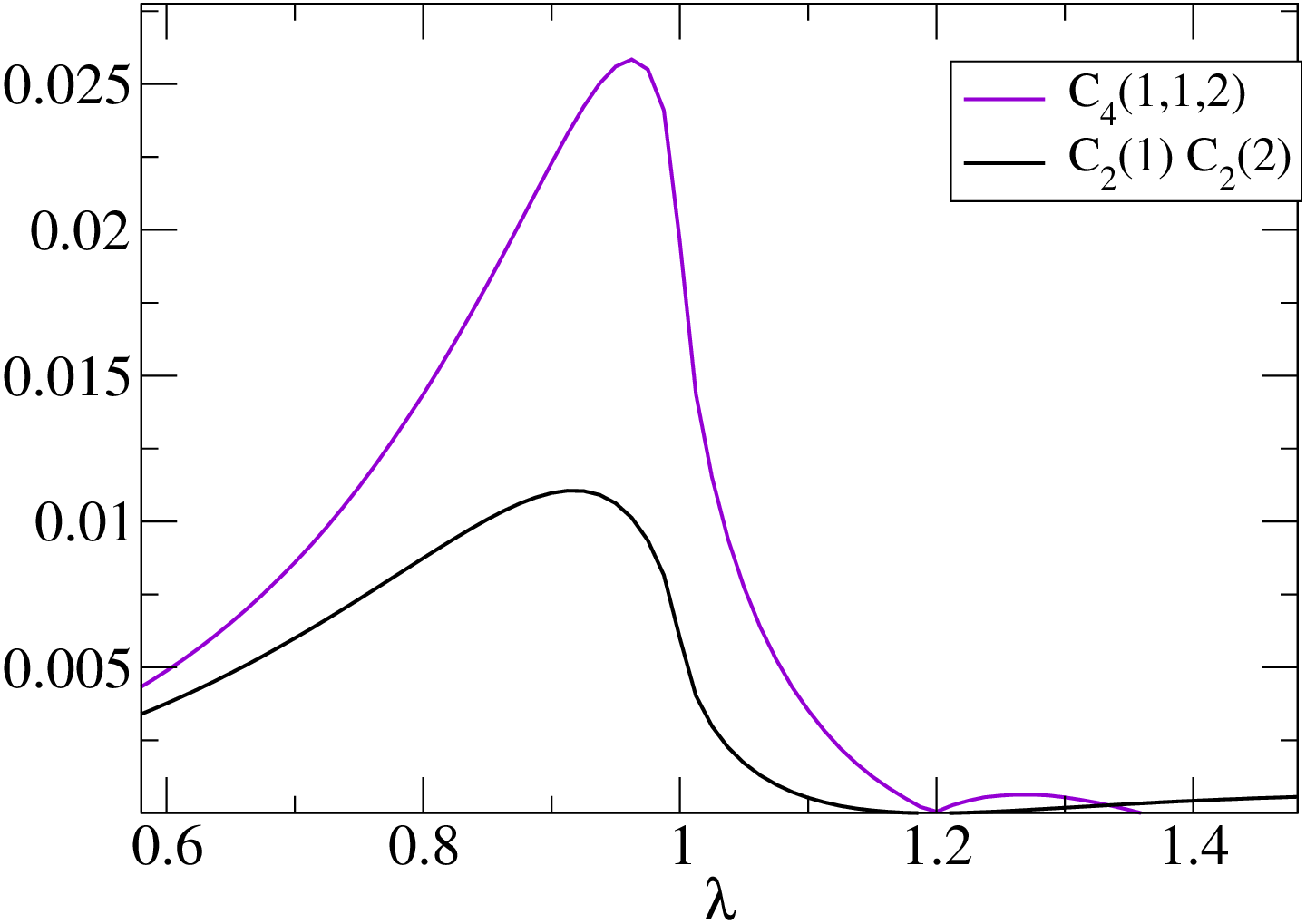}
  \caption{The behavior of $C_4(1,1,2)$ is shown (blue curve) 
and compared with $C_2(1)C_2(2)$ (black curve). The curves are for $\gamma=0.55$.
A similar behavior is observed as for $\gamma=0.58$ and $\gamma=0.59$ 
(see figs. \ref{C4-112-gamma0.58} and \ref{C4-112-gamma0.59}, respectively.}
  \label{C4-112-gamma0.55}
\end{figure}

\begin{figure}
\centering
  \includegraphics[width=.8\linewidth]{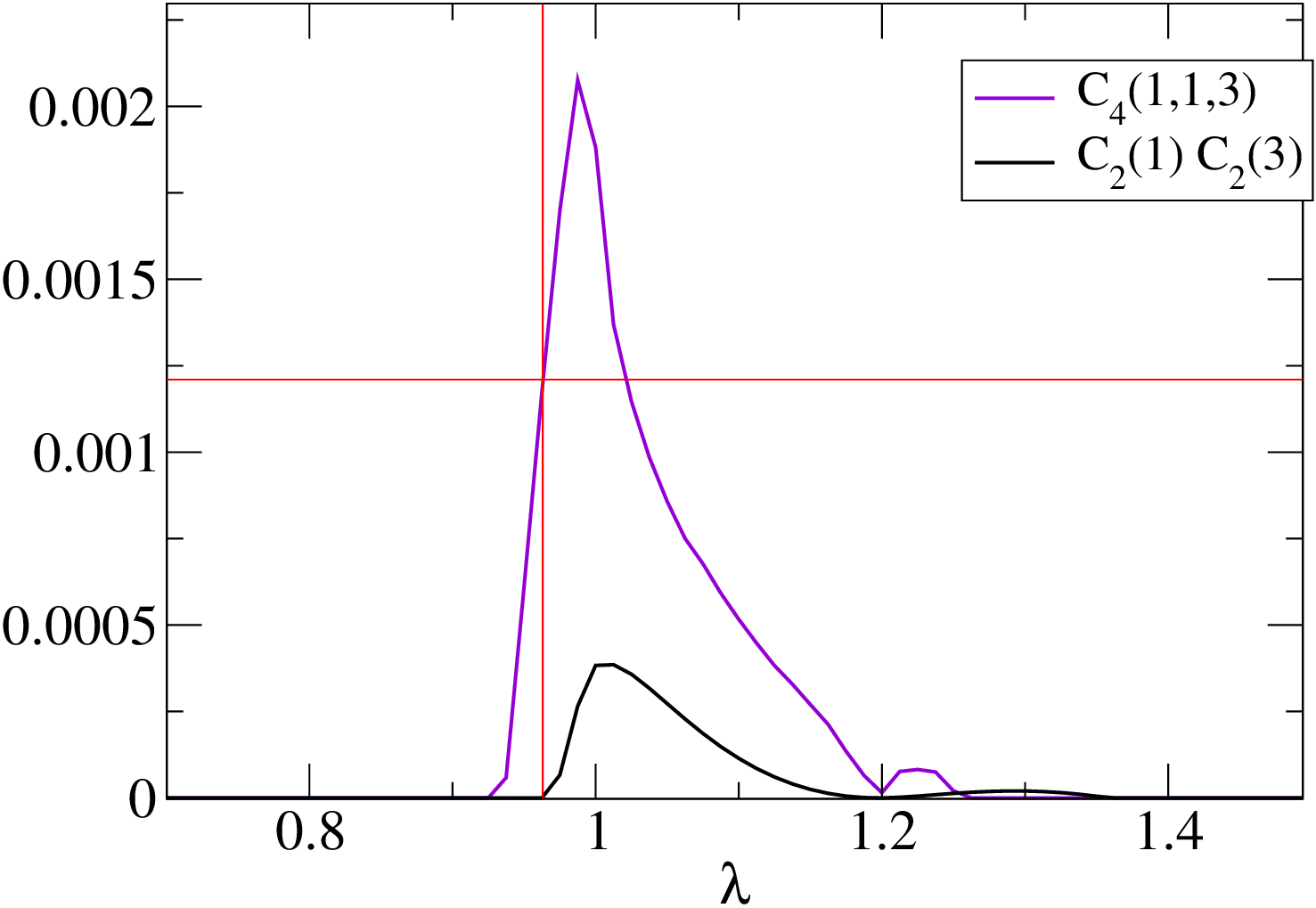}
  \caption{$C4(1,1,3)$ already sets in considerably earlier than $C_2(1)C_2(3)$. 
The value of $C_4(1,1,3)$ where $C_2(1)C_2(3)$ sets in is considerably above $0.0012$,
which is more than $50\%$ of its maximum value. }
  \label{C4-113-gamma0.55}
\end{figure}

\begin{figure}
\centering
  \includegraphics[width=.45\textwidth]{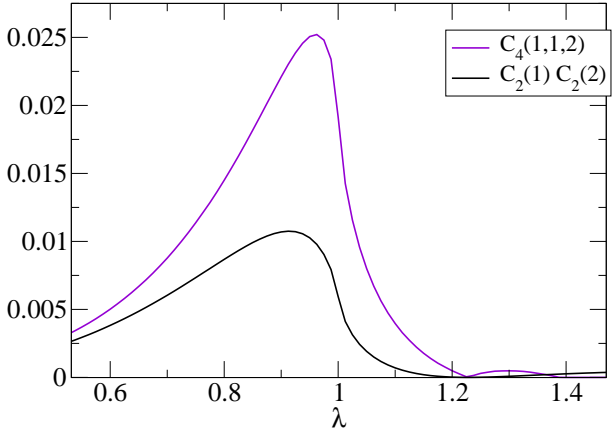}
  \caption{$C_4(1,1,2)$ together with $C_2(1)C_2(2)$ is shown for $\gamma=0.58$. The plot is basically
as in fig. \ref{C4-112-gamma0.55}, except that the factorizing point has moved as
$\lambda_f=(1-\gamma^2)^{-\frac{1}{2}}$.}
  \label{C4-112-gamma0.58}
\end{figure}

\begin{figure}
\centering
  \includegraphics[width=.45\textwidth]{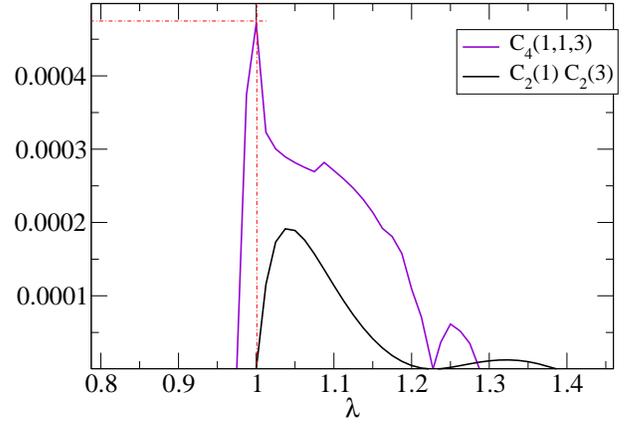}
  \caption{$C_4(1,1,3)$ together with $C_2(1)C_2(3)$ is shown for $\gamma=0.58$. Whereas $C_4(1,1,3)$ 
has squeezed apparently against the critical point, $\lambda_c$, with a high, apparently infinite, 
slope. It appears to destructively 
interfere with something that has approximately the same height as $C_2(1)C_2(3)$.}
  \label{C4-113-gamma0.58}
\end{figure}

\begin{figure}
\centering
  \includegraphics[width=.75\linewidth]{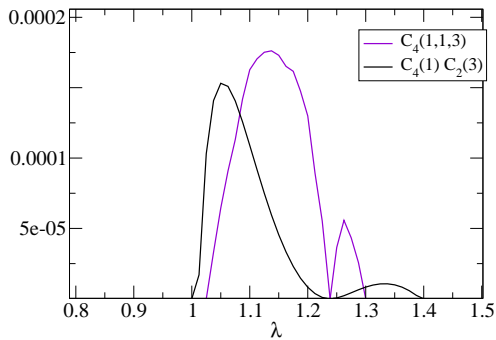}
  \caption{$C_4(1,1,3)$ together with $C_2(1)C_2(3)$ is shown for $\gamma=0.59$. It has ``overtaken''
$C_2(1)C_2(3)$, which is still pinned at $\lambda_c$.}
  \label{C4-113-gamma0.59}
\end{figure}

Next we look at $1-1-n$ configurations.
This state should become a tensor product for growing number of $n$. Hence, its
4-tangle tends to zero.
We analyze the 4-tangle $C_4(1,1,n)$ for different values of the anisotropy parameter $\gamma$
and for $n=2$ and $3$. We observe that $C_4(1,1,2)$ doesn't differ much for the values of $\gamma$ from $0.55$
via $0.58$ to $0.59$ (figs. \ref{C4-112-gamma0.55}, \ref{C4-112-gamma0.58}, and \ref{C4-112-gamma0.59})
besides the shift of the factorizing point following $\lambda_f=(1-\gamma^2)^{-1/2}$.
The interval of $\gamma$ is chosen such that $C_2(3)$, at
the critical value, drops to zero a bit before $\gamma=0.58$. 
Something interesting begins to happen, when the 4-tangle of the distance $1-1-3$ is considered.
Whereas for $\gamma=0.55$, $C_4(1,1,3)$ sets in considerably before $C_2(1)C_2(3)$, 
$C_2(1)C_2(3)$ begins to have non-vanishing values from about $\lambda=0.975$, with a
visible finite slope, 
a bit before the critical point $\lambda_c=1$ (see fig. \ref{C4-113-gamma0.55}).
Then, $C_2(1)C_2(3)$ behaves as if it were ``pinned'' at the critical point $\lambda_c$ 
for the following two values 
of $\gamma=0.58$ (both curves with a high, quasi infinite, slope; see fig. \ref{C4-113-gamma0.58}) 
and also $\gamma=0.59$ (again with a visible slope; see figure~\ref{C4-113-gamma0.59}).
$C_4(1,1,3)$ is definitely feeling the critical point as well: whereas its onsetting 
remains at about the same distance from the point where $C_2(1)C_2(3)$ sets in from $\gamma=0.55$
to $\gamma=0.58$ it however squeezes the function $C_2(1)C_2(3)$ against the critical point,
and therby also feels an apparently destructively interfering part from it 
(see the maximum of $C_4(1,1,3)$ in fig. \ref{C4-113-gamma0.58}). 
At $\gamma=0.59$, $C_4(1,1,3)$ has already overtaken $C_2(1)C_2(2)$, the latter being 
still stuck to the critical $\lambda_c$ but with a visible slope.

We remember that a non-vanishing $C_4$ in presence of a zero $C_2(1)C_2(2)$ 
doesn't need to mean a non-zero portion of $GHZ$-like entanglement in principle 
(see the discussion of fig. \ref{C4-W-BellBells}), 
it is however an interesting observation, 
which would certify $GHZ$ entanglement, if the genuine negativity would be available.
It would not exclude 
genuine multipartite entanglement to be there, since there 
are states that are genuine multipartite entangled states 
(e.g. cluster states and X-states\cite{OS04,OS05,OS09})
for which the 4-tangle $C_4$ vanishes. 

\begin{figure}
\centering
  \includegraphics[width=.75\linewidth]{C4112C1C2.gamma.59.eps}
  \caption{$C_4(1,1,2)$ together with $C_2(1)C_2(2)$ is shown for $\gamma=0.59$. The plot is basically
as in fig. \ref{C4-112-gamma0.55}, except that the factorizing point has moved as
$\lambda_f=(1-\gamma^2)^{-\frac{1}{2}}$.}
  \label{C4-112-gamma0.59}
\end{figure}

We have to mention that the function $C_2(3)$, as every entanglement measure,
is pinned at and also localized about the factorizing field. Therefore it vanishes 
for the Ising model\cite{OstNat}. It exists for an arbitrary value of $0<\gamma<1$ and
will be accompanied by $C_4(1,1,3)$ (also pinned at the factorizing field), 
getting smaller and smaller as $\gamma\to 1$. Similar conclusions apply also to $C_2(n)$
and we conjecture also the corresponding behavior for $\gamma\to1$ of $C_4(1,1,n)$.

\section{Conclusions and outlook} \label{concls}

We have analyzed the 4-tangle $C_4$ for the transverse Ising and transverse XY model. $C_4$ only measures two 
types of entanglement, the GHZ-type and the tensor product of Bell states since it is an invariant polynomial 
measure of degree two and only the GHZ state and the product of Bell states have a minimal irreducible length in 
their representation which is two\cite{OS09} (Bell and in general the GHZ states can be written as a superposition 
of two orthogonal product basis states). We have analyzed the 4-tangle in some detail in chapter~\ref{4-tangle}
and conclude that there is no simple way extracting whether GHZ-entanglement is in the state or a product of Bell 
states from $C_4$ alone; some further measure of entanglement must be analysed as well.
We only have a hint towards GHZ-entanglement in that both the genuine negativity and $C_4$ 
have the same order of magnitude and they behave the same way.
A possibility would consist in three particle measures of entanglement like the threetangle or 
the three further measures of genuine four-particle entanglement.
One such extension exist for three particles\cite{Siewert2012,Eltschka2012} and should be evaluated.
However, for the four particle case one first has to give an elaborate extension to mixed states. 

We then analyze the transverse Ising model and more in general the tranverse XY model. For the Ising model, 
$C_4$ satisfies the necessary requirement of being upper bound by the product of the concurrences 
in the example $C_4(1,n,1)$ for $n\geq 3$ in case of a product state $\rho_2\otimes\rho_2$.
It is for $n\geq 7$ exactly zero close to the critical point 
(we didn't check this for $n>12$ but formulate it as a surmise here).
For $C_4(2,n,2)$ this requirement is trivially satisfied in that $C_4(2,n,2)$ vanishes exactly for every value
$\gamma$ and $\lambda$.
For configurations of distances $1-n-2$ only $C_4(1,1,2)$ (or eqivalently $C_4(2,1,1)$) 
gives a non-zero result. For $n=1$ the state is not
to a good approximation a product state and hence $C_4(1,1,2)$ is not bound by $C_2(1)C_2(2)$. 
For the case of states with distances $(i,n,j)$ and $n\to\infty$ one is roughly left with a tensor product 
of density matrices and there the 4-tangle $C_4$ has to be upper bounded by the product of the 
concurrences $C_2(i)C_2(j)$.
Also here, a hint is given that the entanglement is of $GHZ$ type 
for the distances $(1,2,1)$ and $(1,1,2)$, 
since the corresponding genuine negativity behaves essentially the same as the genuine negativity\cite{Hofmann14}.
However, additional entanglement measures are mandatory in order to get a clear answer.

The transverse XY model behaves essentially the same way. $C_4(1,n,1)$ is upper bounded by the 
concurrence $C_2(1)^2$ for $n\geq 4$ and $\lambda\lesssim 1.25$ and vanishes precisely for $n\geq 7$ close 
to the critical point. At the factorizing point, this doesn't appear to be the case.
There, it doesn't satisfy the necessary condition for a tensor product for $n<8$ and 
$1.25<\lambda <\lambda_f$. For $\lambda>\lambda_f$ this condition is satisfied. 
It becomes particularly interesting when we analyse $C_4(1,1,n)$ for $n=2,3$. Here an interesting 
interference phenomenon occurs, which needs further analysis. 

\section*{Acknowledgements}

We acknowledge financial support by the German Research Foundation within the SFB TR12.


\end{document}